\documentclass{emulateapj}

\usepackage{epsfig}

\newcommand{\hMpc}{\ensuremath{\,h^{-1}\,{\rm Mpc}}}

\newcommand{\avg}[1]{\ensuremath{\langle #1 \rangle}}

\begin{document}

%%%%%%%%%%%%%%%%%%%%%%%%%%%%%%%%%%%%%%%%%%%%%%%%%%%%%%%%%%%%%%%%%%%%%%%%%%%
% Front Material
%%%%%%%%%%%%%%%%%%%%%%%%%%%%%%%%%%%%%%%%%%%%%%%%%%%%%%%%%%%%%%%%%%%%%%%%%%%

%\twocolumn[%%% Begin front material

%\journalid{337}{15 January 1989}
%\articleid{11}{14}
 
\submitted{\today. To be submitted to \apj.}

%\title{The Mean Absorption in the Ly$\alpha$ Forest:
%Sightline-to-Sightline Scatter in the Post-reionization Universe}
\title{Have We Detected Patchy Reionization in Quasar Spectra?}
\author{Adam Lidz$^{1}$, S. Peng Oh$^{2}$, and Steven
R. Furlanetto$^{3}$}   \affil{$^1$ Harvard-Smithsonian Center for
Astrophysics, 60 Garden Street, Cambridge, MA 02138, USA\\ $^2$
Department of Physics, University of California, Santa Barbara, CA
93106\\ $^3$ Division of Physics, Mathematics, \&
Astronomy, California Institute of Technology, Mail Code 130-33, Pasadena,
CA 91125\\ Electronic mail : {\tt alidz@cfa.harvard.edu,
peng@physics.ucsb.edu, sfurlane@tapir.caltech.edu}}
%\email{emails}

\begin{abstract}
The Ly$\alpha$ forest at $z \gtrsim 5.5$  shows strong scatter in the mean
transmission even when smoothed over very large spatial scales, $\gtrsim 50$
Mpc/$h$. This has been interpreted as a signature of strongly
fluctuating radiation fields, or patchy reionization. To test this claim, we
calculate the scatter arising solely from density fluctuations, with a
uniform ionizing background, using analytic arguments and simulations.
This scatter alone is comparable to that observed. It rises steeply with redshift and is
of order unity by $z \sim 6$, even on $\sim 50$ Mpc/$h$ scales.
This arises because:
i) at $z \sim 6$, transmission spectra, which are sensitive mainly to rare voids, are 
highly biased (with a linear bias factor $b \geq 4-5$) tracers of underlying density 
fluctuations, and ii) projected power from small-scale transverse modes is aliased to long 
wavelength line-of-sight modes. Inferring patchy reionization from quasar spectra is therefore subtle and 
requires much more detailed modeling. Similarly, we expect order unity transmission
fluctuations in the $z \sim 3$ HeII Ly$\alpha$ forest from density fluctuations
alone, on the scales over which these measurements are typically made.
\end{abstract}

%]%%% End front material

%%%%%%%%%%%%%%%%%%%%%%%%%%%%%%%%%%%%%%%%%%%%%%%%%%%%%%%%%%%%%%%%%%%%%%%%%%%
\section{Introduction} \label{intro}
%%%%%%%%%%%%%%%%%%%%%%%%%%%%%%%%%%%%%%%%%%%%%%%%%%%%%%%%%%%%%%%%%%%%%%%%%%%

At redshifts close to $z \sim 3$, the structure in the Ly$\alpha$ forest has been shown to arise
naturally from density fluctuations in the cosmic web (e.g. Miralda-Escud\'e et al. 1996). 
At sufficiently high redshift, however, the structure in 
the Ly$\alpha$ forest may instead 
largely reflect the topology of reionization, and/or a strongly fluctuating radiation field.
In high-redshift quasar spectra with extended opaque regions, significant gaps of 
substantial transmission occur
(e.g. Becker et al. 2001, White et al. 2003, White et al. 2005). This has previously been attributed to 
a strongly fluctuating UV background, as expected at the tail end of 
reionization (Wyithe \& Loeb 2005, Fan et al 2005). 

Could these transmission gaps simply arise from underdense regions where the neutral 
hydrogen fraction is lower? The transmission in the $z \sim 6$ quasar spectra differs
significantly from sightline to sightline, even when one averages 
over co-moving length scales 
of $\sim 50 - 100$ Mpc/$h$. Since the density variance is small over such large scales, one 
might naively expect that the 
reionization of the IGM must be {incomplete} near $z \sim 6$.

In this {\it Letter}, we critically examine this naive intuition. Is rapidly increasing scatter in 
sightline to sightline flux transmission a good diagnostic for patchy 
reionization (Fan et al. 2002, Lidz et al. 2002, Sokasian et al. 2003, Paschos \& Norman 2005)? 
%How large should the sightline to sightline scatter
%in the mean transmission bethe {\em null hypothesis} that all of the 
%structure in the $z \sim 6$ Ly$\alpha$ forest is the result of density fluctuations?
%The basic quantity we calculate, the sightline to sightline scatter in the mean transmission,
%has been considered previously by other authors at low redshift (Zuo 1993, Tytler et al. 2004),
%and at high redshift, as a diagnostic for patchy reionization (Fan et al. 2002, Lidz et al. 2002,
%Sokasian et al. 2003, Paschos \& Norman 2005). Presently, w
We find that in fact the fractional scatter in the mean transmissivity of the IGM will be large 
at high redshift, even for a completely uniform ionizing background. An analogous calculation 
applies to the case of the HeII Ly$\alpha$ forest near $z \sim 3$.

%%%%%%%%%%%%%%%%%%%%%%%%%%%%%%%%%%%%%%%%%%%%%%%%%%%%%%%%%%%%%%%%%%%%%%%%%%%
\section{The Flux Power Spectrum}
\label{model}
%%%%%%%%%%%%%%%%%%%%%%%%%%%%%%%%%%%%%%%%%%%%%%%%%%%%%%%%%%%%%%%%%%%%%%%%%%%

We adopt the usual `gravitational instability' model of the Ly$\alpha$ forest 
valid at $z\sim 3$ (e.g., Hui et al 1997). In particular, we assume an isothermal gas and a 
uniform radiation field, to see if these assumptions demonstrably break down at $z\sim 6$. 
Assuming 
photoionization equilibrium, the Ly$\alpha$ optical depth is $\tau = A \Delta^{2}$, 
where $A \propto (1+z)^{4.5}T^{-0.7}/\Gamma$, and the transmitted flux is $F=e^{-\tau}$. We study 
the 
fluctuations in transmitted flux, $\delta_F = ({F - \avg{F}})/{\avg{F}}$, and the line of sight 
power spectrum of these fluctuations, $P_{\rm F}(k)$. We will also refer to the effective 
optical depth, $\tau_{\rm eff} = -{\rm ln} \avg{F}$.
 
It is important to note that $P_{\rm F}(k)$ is related to the underlying power spectrum of 
density fluctuations $P_{\delta}(k)$ in a rather complicated way. 
First, gas pressure smooths the baryon-density distribution on small scales with respect to the
dark matter. Second, a non-linear transformation maps density into optical depth ($\tau \propto \Delta^{2}$).
Third, one maps from real space to skewers in redshift space, projecting from 3D to 1D and incorporating the
effect of peculiar velocities. Fourth,
one convolves the optical depth distribution with the thermal broadening kernel. Finally, a second non-linear 
transformation maps $\tau$ into transmitted flux ($F= e^{-\tau})$.
These issues are well
studied in the context of the $z \sim 3$ Ly$\alpha$ forest (e.g. Croft et al. 2002) but are 
under-appreciated in
the reionization literature. Previous neglect of this physics has led to erroneous conclusions. 
%It is essential to model this physics in order to reliably
%estimate transmissivity fluctuations at high redshift.

\subsection{Analytic Estimates}
\label{estimates}

We will therefore require numerical simulations to model the flux power spectrum in detail, and 
its 
dependence on redshift and the mean transmissivity of the IGM. We can nonetheless anticipate the results with
analytic arguments. Specifically, we estimate the point-to-point flux variance as a 
function
of $A$ and $\tau_{\rm eff}$. We perform this calculation using the
Miralda-Escud\'e et al. (2001) fitting formula for the gas density PDF: 
%(following e.g., Fan et al. 2002, Songaila \& Cowie 2002):
\begin{equation}
P(\Delta) d\Delta = G \Delta^{-b} \rm{exp} \left[- \frac{\left(\Delta^{-2/3} - {\it C} \right)^2}{8 \delta_0^2/9} \right] {\it d}\Delta.
\label{pdf_miralda}
\end{equation}
Here the parameters $G$ and $C$ are fixed by requiring the PDF to
normalize to unity and to satisfy $\langle \Delta \rangle = 1$. The other
parameters are given by $\delta_0=7.61/(1+z)$ and $b=2.5$ at $z=6$. 
We then calculate the first two moments of the (unsmoothed) flux distribution,
$\avg{F} = \avg{e^{-A \Delta^2}}$, and $\avg{F^2} = \avg{e^{-2 A \Delta^2}}$, by integrating 
over the gas density PDF. These estimates ignore the effect of peculiar velocities and thermal
broadening, but we include these effects subsequently using our simulations.

These integrals can be approximated using the method of steepest 
descents (Songaila \& Cowie 2002), to 
give $\avg{F} \sim c \tilde{A}^{1/4} {\rm exp}[-d \tilde{A}^{0.4}]$, and 
$\avg{F^2}/\avg{F}^2 \sim \tilde{c} \tilde{A}^{-1/4} {\rm exp}[\tilde{d} \tilde{A}^{0.4}]$, 
where $c$, $d$, $\tilde{c}$, and $\tilde{d}$ are constants, and $\tilde{A}=A/25$. We find 
that $c = 5.3$, $d = 5.1$,
$\tilde{c} = 0.22$, and $\tilde{d} = 3.5$ provide a good approximation to the results of
full numerical integrations. These scalings immediately suggest that the fractional 
point-to-point dispersion will be quite large when $A$ and $\tau_{\rm eff}$ are also large,
{\em even assuming a homogeneous radiation field.}
For instance, when $\tau_{\rm eff} \sim 5$, we expect the fractional point-to-point dispersion to
be $\sigma_F/\avg{F} \sim 4$. On the other hand, when $\tau_{\rm eff} \sim 1$, the point-to-point
dispersion is only $\sigma_F/\avg{F} \sim 0.8$. When the transmission becomes small, it is dominated by rare voids, which drastically increases the point-to-point variance. 

\subsection{Numerical Calculations}
\label{sims}

In order to make a more accurate calculation, and to consider transmission fluctuations smoothed over 
large scales, we 
calculate the flux power spectrum numerically. We use
a Hydro-Particle-Mesh (HPM) (Gnedin \& Hui 1998) simulation with $2 \times 512^3$
particles and $512^3$ mesh-points in a $40 \hMpc$ box, assuming a $\Lambda$CDM cosmology 
with $(\sigma_8,h,\Omega_{b} h^{2},\Omega_{m},\Omega_{\Lambda})=(0.84,0.7,0.02,0.3,0.7)$. The 
simulation is 
an implementation of HPM into the parallel N-body code, Mc$^{2}$ (see Heitmann et al. 2005 for detailed 
tests of Mc$^{2}$, and the Appendix of Lidz et al. 2005, Habib et al. 2006, in prep., for the details and 
convergence studies of our implementation of HPM).

We take the baryon density and peculiar velocity fields from the simulation and
extract artificial spectra in the usual manner (for more details see e.g., Hui, Gnedin \& Zhang 1997). 
Specifically, 
we consider $z = 4.9, 5.3, 5.7, 6.0$, and adjust $A$ to match 
$\avg{F} = 0.20, 0.13, 0.06$, and $0.01$. These values are close to, but slightly larger than,
the observational measurements of Becker et al. (2001), White et al. (2003), and Fan et al. (2005). 
For each redshift, we extract $5,000$ artificial spectra from the simulation
box.

\begin{figure}[t]
%\plotone{pauto_z6_mf0.01.ps}
\vbox{ \centerline{ \epsfig{file=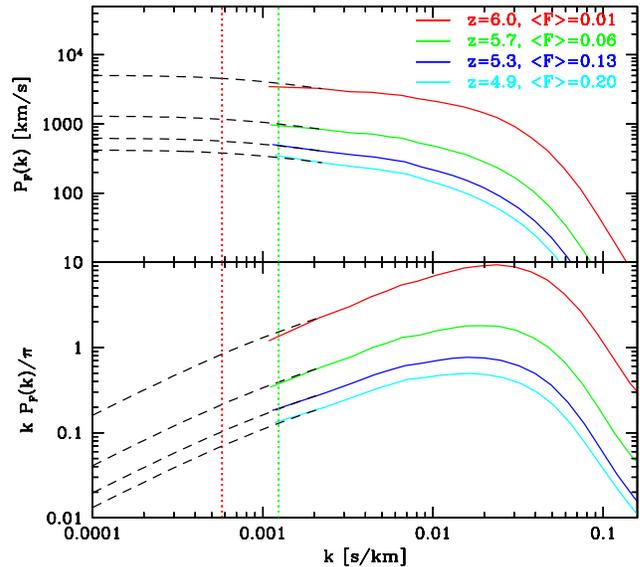,width=9.5truecm}}
\caption{\footnotesize Flux power spectrum as a function of scale at various 
redshifts, with $A$ adjusted 
to the mean flux level $\langle F \rangle$ shown. The black dashed lines show the large-scale 
extrapolations of 
the simulation results described in the text.
The green and red dotted lines correspond to the length scales $L = 35,75$ Mpc/$h$ at $z =
6$ respectively. Notice that $k P_{\rm F} (k)$ decreases only slowly towards
large scales, a consequence of aliasing, as we describe in the
text. At $z \sim 6$, the fluctuations are of order unity
even on scales close to $\sim 50$ Mpc/$h$.}
\label{pauto}}
\end{figure}

In order to consider large-scale transmission fluctuations, we need to extrapolate to scales 
beyond the size of the simulation box. We do so assuming that on large scales, the 
flux power spectrum and the 1D linear matter power spectrum differ only in normalization and 
not in shape (linear biasing, e.g., Scherrer \& Weinberg 1998, Lidz et al. 2002), although we
will show in \S \ref{scatter} that the bias {\em increases sharply with increasing redshift}. We 
normalize the extrapolation to the 
simulated flux power spectrum at $k \sim k_{f}/2$, 
where $k_{f}$ is the wavenumber associated with the fundamental mode of the box. We 
generally consider
smoothing scales only somewhat larger than the size of our 
simulation box, $L_{\rm box} = 40$ Mpc/$h$. 
Furthermore, the flux power spectrum is relatively flat on these scales, and so our results 
are not
terribly sensitive to the details of this extrapolation.\footnote{For instance, if we 
extrapolate our results using
a power law, our estimate of the transmission dispersion, Eq. (\ref{var_mf}), changes 
by $\sim 20\%$.} 

In Fig \ref{pauto}, we show the flux power spectrum from our simulations. The amplitude of transmission fluctuations 
clearly increases rapidly with redshift, despite declining density fluctuations. In fact, 
fractional fluctuations 
are of {order unity} on scales
of $L \sim 35$ Mpc/$h$ at $z \sim 6$, {\it even for a uniform radiation field}. This implies 
that it may be 
difficult to discern patchy reionization from transmission fluctuations. We study this further 
in \S \ref{scatter}. 

Another interesting feature of Fig \ref{pauto} is that the flux power spectrum is quite 
flat: $k P_F(k)$ declines only gradually towards large scales. This is because absorption 
spectra 
provide only a 1D skewer through the IGM, with a power spectrum (ignoring redshift space distortions for 
illustrative purposes):
\begin{equation} \label{alias_pow}
P_{\rm 1d}(k_\parallel) = \int_{k_\parallel}^{\infty} \frac{dk}{2\pi} k P_{\rm 3d} (k).
\end{equation}
Thus, the 1D power spectrum falls off much more slowly towards large scales than 
the 3D power spectrum. This is the well-known process of aliasing found in pencil-beam 
galaxy surveys (Kaiser \& Peacock 1991), whereby short transverse wavelength modes in projection alias 
power to long wavelength, line of sight modes. To illustrate this quantitatively, we consider the
amplitude of 3D fluctuations in the non-linear baryon-density field, 
$\Delta^2_b(k) = k^3 P_b(k)/(2 \pi^2)$, approximating the relation between the baryonic and dark
matter power spectra as $P_b(k) = e^{-2 k^2/k_f^2} P_{\rm dm} (k)$, assuming
$k_f = 20 h$ Mpc$^{-1}$ (Gnedin \& Hui 1998), and using fitting formulae from
Peacock \& Dodds (1996) and Eisenstein \& Hu (1999). We contrast this with the variance 
of 1D baryonic
fluctuations, $\Delta^2_{\rm 1D, b}(k_\parallel) = k_\parallel P_{\rm 1D, b}(k_\parallel)/\pi$,
comparing line of sight fluctuations of wavenumber $k_\parallel$ to 3D fluctuations with 
wavenumber
$|\vec{k}| = k_\parallel$. The 1D power spectrum is a factor of $\sim 5.5 - 10.6$ times 
larger than
the 3D power spectrum for modes with wavelength $L \sim 50 - 100$ Mpc/$h$. Fluctuations averaged
over a skewer of a given length are much larger than fluctuations averaged over 
an entire sphere of comparable size.

\section{The Sightline to Sightline Scatter}
\label{scatter}

Given $P_F(k)$, the fractional sightline-to-sightline variance in the 
transmission is (Lidz et al. 2002):
\begin{equation} \label{var_mf}
\frac{\sigma^2_{\rm \avg{F}}}{\avg{F}^2} = 2 \int_{0}^{\infty} \frac{dk}{2 \pi} 
P_F(k) \left[ \frac{{\rm sin}(k L/2)}{(k L/2)} \right]^{2}.
\end{equation}
Shot-noise is negligible for significant stretches
of spectrum with at least moderate signal to noise (see Lidz et al. 2002 for details). 
%The term in brackets, represents the smoothing window used to define the mean transmission, 
%which we take to be a top-hat in real space.
We consider the fractional dispersion in the mean transmitted
flux, $\sigma_{\avg{F}}/\avg{F}$, as opposed to the absolute 
dispersion, $\sigma_{\avg{F}}$. Note that $\sigma_{\tau_{\rm eff}}=\sigma_{\avg{F}}/\avg{F}$ for
small fluctuations.
%The effective optical depth, $\tau_{\rm eff}$, is generally plotted in the literature.
%This quantity depends logarithmically on the transmission, and so the fractional dispersion
%is the more relevant quantity.

\begin{figure}[t]
\vbox{ \centerline{ \epsfig{file=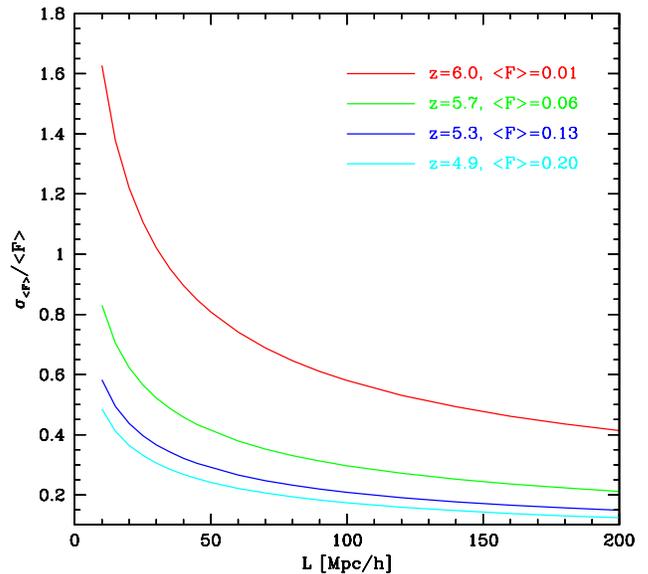,width=9.5truecm}}
%/home/alidz/FLUX_z6_new/FIT_flux/sigma_v_smooth_new.eps
\caption{\footnotesize Sightline-to-sightline scatter in the mean transmission as a 
function of scale 
and redshift. The cyan, blue, green, and red curves indicate the expected scatter in the mean
transmission as a function of scale at redshifts $z = 4.9, 5.3, 5.7$ and $6$ respectively.
The fractional scatter increases rapidly with decreasing transmission, and dies off only
slowly towards large scales.}
\label{scatter_smooth}}
\end{figure}

%A calculation of the sightline to sightline dispersion in the transmission for a range of redshifts and
%opacities is shown 
We show this quantity in Fig. \ref{scatter_smooth}. 
%The qualitative trends seen in the flux 
%power spectrum 
%calculation (Fig. \ref{pauto}) are naturally mirrored in the calculation of the flux %variance:
As in the flux 
power spectrum 
calculation (Fig. \ref{pauto}), the fractional dispersion increases rapidly with decreasing transmission, and 
falls off only slowly as the smoothing scale becomes large\footnote{Since, on large scales, the flat $P_{F}(k)$ 
is close to a white noise power spectrum, $\sigma^2_{\rm \avg{F}}/\avg{F}^{2} \propto L^{-1}$.}. By $z \sim 6$, 
assuming $\avg{F} = 0.01$ ($\tau_{\rm eff} =4.6$), a complete absorption
trough $\langle F \rangle =0$ of length $\sim 50$ Mpc/$h$ is only a $\sim 1-\sigma$ 
fluctuation, even though density fluctuations are small, $\sigma_{\rho, 1D} \sim 0.2$ on these scales. 
Transmission fluctuations are a {\rm highly biased} tracer of underlying density fluctuations
at high redshift, with a bias factor 
of $b = (\sigma_{\avg{F}}/\avg{F})/\sigma_{\rho, 1D} \sim 4$ when $\tau_{\rm eff} \sim 4-5$ (and 
increasing at higher $\tau_{\rm eff}$). 
Although on large scales the flux and density power spectra have the same shape, one can {\em not} then
simply assume that $\sigma_{\avg{F}}/\avg{F} \sim \sigma_{\rho, 1D}$, as has previously been done (Wyithe \& Loeb 2005,
Fan et al. 2005).

Note that state of the art reionization simulations have thus far been done
with boxsizes of less than $10$ Mpc/$h$ (e.g. Gnedin 2000, Sokasian et al. 2003, Paschos \& Norman 2005), although
steps towards large scale simulations are being made (Kohler et al. 2005, Iliev et al. 2005).
It is common practice to generate mock spectra from these simulations by 
wrapping long lines of sight around these small periodic boxes many times. 
From Fig. \ref{pauto} and Eq. \ref{var_mf}, it is 
clear that estimates of the transmissivity scatter from mock spectra generated in this way are 
{\em underestimates}. For instance, the best case scenario for extrapolating Paschos \& Norman (2005)
to larger scales would have a flat line of sight power spectrum on scales larger than their
box ($L_{\rm box} = 6.8$ Mpc/$h$) and match our results on smaller scales. Even in this best case,
we calculate that they underestimate the transmissivity
variance by a factor of $\sim 2$ for a smoothing scale of $L = 50$ Mpc/$h$.

\begin{figure}[t]
\vbox{ \centerline{ \epsfig{file=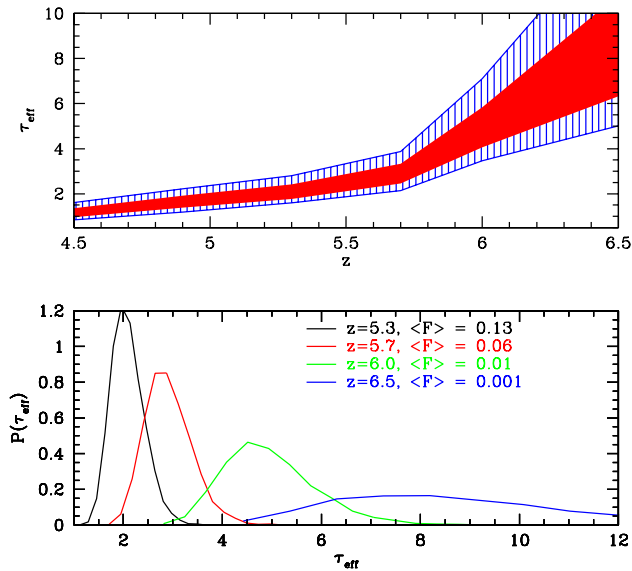,width=9.5truecm}}
%/home/alidz/FLUX_z6_new/PDF_los/pdf_tau_variance_new.eps
\caption{\footnotesize Evolution of $\tau_{\rm eff}$ and the expected sightline to sightline 
scatter. The top panel shows $68\%$ and $95\%$ confidence regions for the effective
optical depth, given a smoothing length of $L = 40$ Mpc/$h$. 
Notice that the shaded regions become quite
broad at large $\tau_{\rm eff}$. We assume central $\tau_{\rm eff}$ values of
$\avg{F} = 0.32, 0.20, 0.13, 0.06, 0.01,$ and $0.001$, at $z = 4.5, 4.9, 5.3, 5.7, 6.0,$
and $6.5$ respectively. The upper limits on $\tau_{\rm eff}$ at $z = 6.5$ extend beyond the edge of the plot.
They are $\tau_{\rm eff} = 10.8, 14$ at $68\%$ and $95\%$ confidence, respectively. The bottom panel 
shows examples of the corresponding PDFs.}
\label{tau_var}}
\end{figure}

Equation (\ref{var_mf}) does not capture the full story since the flux probability
distribution is {non-Gaussian}. This is obvious since fluctuations in the 
mean transmission are of order unity, yet the transmission is bounded between zero and one. 
To reliably calculate confidence intervals, we thus need to consider the full shape of the PDF
of the transmission, smoothed on large scales.
Because the extrapolation of the full PDF to large scales is 
non-trivial (unlike for the simple power spectrum), we will calculate the PDF directly from the simulation
box. Thus we effectively smooth on scales of $L=L_{\rm box}=40$ Mpc/$h$ but underestimate the width of
the PDF because of large-scale power missing from our box. We show the PDF of $\tau_{\rm eff}$ in the bottom 
panel of Fig. \ref{tau_var}. It becomes increasingly broad and skewed,
with a long tail towards high optical depth at high redshift. This asymmetry implies a bias
in measurements of $\tau_{\rm eff}$ from samples with a small number of sightlines.
%The bottom panel of the figure further illustrates that, even adopting a large smoothing scale, 
%the scatter increases rapidly with redshift and will be quite large near $z \sim 6$. We 
%therefore caution against over-interpreting the increasing scatter seen in plots 
%of $\tau_{\rm eff}(z)$ in the literature.

The top panel of Fig. \ref{tau_var} shows error bands 
for the expected sightline
to sightline scatter in the effective optical depth as a function of redshift, adopting a smoothing length of $L = 40$ Mpc/$h$. Even for large smoothing scales, the scatter increases rapidly with redshift. We therefore caution against over-interpreting the increasing scatter seen in plots of $\tau_{\rm eff}(z)$ seen in the literature

How does the level of scatter we predict compare with existing observations?
We compare with recent measurements from Fan et al. (2005) (their Table 5), which indicate
a dispersion of $\tau = (2.1 \pm 0.3, 2.5 \pm 0.5, 2.6 \pm 0.6, 3.2 \pm 0.8, 4.0 \pm 0.8, 7.1 \pm 2.1)$ at
$z = (4.90 - 5.15, 5.15 - 5.35, 5.35 - 5.55, 5.55 - 5.75, 5.75 - 5.95, 5.95 - 6.25)$. The values
of the dispersion in the last two-redshift bins are lower limits, since some lines of sight
in these bins show complete absorption troughs. Interpolating from our results, which assume a different
redshift evolution for $\tau_{\rm eff}$, we find $\tau = (2.1^{+0.38}_{-0.28}\;{}^{+0.79}_{-0.47}$,
$2.5^{+0.46}_{-0.32}\;{}^{+0.94}_{-0.58}$, $2.6^{+0.48}_{-0.33}\;{}^{+0.98}_{-0.61}$,
$3.2^{+0.67}_{-0.39}\;{}^{+1.4}_{-0.78}$, $4.0^{+0.99}_{-0.48}\;{}^{+2.0}_{-0.98}$, 
$7.1^{+3.9}_{-0.60}\;{}^{+7.1}_{-1.9})$, where the intervals indicate $68\%$ and $95\%$ confidence regions. 
Note that Fan et al. (2005) considered a 
smoothing scale of $\Delta z = 0.15$, corresponding to $L = 44 \ (56)$ Mpc/$h$ at $z= 6 \ (5)$.
Our numbers assume a slightly different scale, $L = 40$ Mpc/$h$, throughout, due to the limited size of
our simulation box.  
From this comparison, we conclude that our results are broadly consistent with the measurements of Fan 
et al. (2005), although we remark that the dispersion is not the ideal statistic to compare with since 
the PDF of $\tau_{\rm eff}$ is asymmetric. We reiterate that {\em our calculations assume
a uniform radiation field and indicate a scatter that is already comparable to the measured values.}

Furthermore, there are several systematic effects, each of which likely tends
to {\em increase} the observed scatter in the mean transmissivity relative to our predictions.
First, as remarked previously, our simulations have a limited boxsize ($L=40$ Mpc/$h$) and therefore
miss large scale power.
Second, at high redshift the quasar continuum is generally estimated using a single power-law, 
extrapolating from redward of the Ly$\alpha$ forest (e.g. Fan et al. 2005). In reality, the 
quasar 
continuum has structure and varies significantly from quasar to 
quasar (e.g. Bernardi et al. 2003).
As a result, some of the apparent transmissivity fluctuations in the data may actually be
due to fluctuations in the quasar continuum.
Next, Lyman limit systems are under-produced in simulations, yet they add large scale 
power to the transmission (e.g. McDonald et al. 2005), which may lead us to underestimate the 
transmissivity scatter. Finally, metal-line 
absorbers are inevitably present in the observed Ly$\alpha$ forest yet are not included
in our calculations. We refer the reader to Tytler et al. (2004) for further quantitative 
discussion of these systematics at $z \sim 2-3$.

\subsection{Sightline-to-Sightline Scatter in the Ly$\beta$ forest}
\label{ly_beta}

How about the transmissivity scatter in the Ly$\beta$ forest? We can estimate 
the fluctuations
in the $z \sim 6$ transmissivity of the Ly$\beta$ region of a quasar 
spectrum using Eq. \ref{var_mf},
%provided we use the power spectrum of fluctuations in the total transmission in this
%region, incorporating fluctuations in the foreground Ly$\alpha$ transmission, as well as 
%fluctuations in the high redshift Ly-$\beta$ transmission. 
but we must consider fluctuations in the foreground Ly$\alpha$ forest as well. 
Dijkstra et al. (2004) show that the total power spectrum $P_{F,{\rm tot.}}$ is:
\begin{eqnarray}
\label{Pkfactorize}
P_{F,{\rm tot.}} (k) &=& P_{F,\alpha} (k) + P_{F,\beta} (k) \\ \nonumber
&+& \int {dk'\over 2\pi} P_{F,\alpha} (k-k') P_{F,\beta} (k').
\end{eqnarray}
%Here $P_{F,{\rm tot.}}$ is the power spectrum
%of the total transmission in the Ly-$\beta$ region, while
%$P_{F,\alpha}$ is the low redshift ($z \sim 4.9$) 
%Ly-$\alpha$ power spectrum 
%and $P_{F,\beta}$ is 
%the high redshift ($z \sim 6$) Ly-$\beta$ power spectrum.
This equation reflects the fact that foreground gas, absorbing in Ly$\alpha$, is widely separated
in physical space, and hence independent of, gas absorbing in Ly$\beta$.
The last term in Eq. \ref{Pkfactorize} occurs because the total correlation function then involves
a product, which transforms into a convolution term. Although higher-order in the transmissivity
fluctuations (Dijkstra et al. 2004), this term cannot be ignored here
because the fluctuations are so large (it increases the total variance by
$\sim 10$--$20 \%$). 

Nonetheless, over an equivalent comoving patch, transmission fluctuations in the Ly$\beta$ forest are 
substantially smaller than in the Ly$\alpha$ forest. We can see this 
heuristically from \S\ref{estimates}: $\avg{F^2}/\avg{F}^2 \sim \tilde{c} \tilde{A}^{-1/4} {\rm exp}[\tilde{d} \tilde{A}^{0.4}]$, 
and $\tilde{A}_{\beta}=\tilde{A}_{\alpha}/6.24$. In practice, this works out to fractional fluctuations which are $\sim 2-4$ times 
smaller in the Ly$\beta$ forest. From the simulations, over a $50 \,{\rm Mpc}\, h^{-1}$ interval, if we 
assume $\langle F_{\alpha} \rangle =(0.2,0.01)$, at $z=(4.9,6)$ (and 
thus $\avg{F_\beta} [z=6] = 0.19$ -- see Oh \& Furlanetto 2005 for details on the relation between 
the Ly$\alpha$ and Ly$\beta$ absorption in a clumpy IGM), we 
obtain $\sigma_{\rm \avg{F_{tot}}}/\avg{F_{\rm tot}} \sim 0.35$, compared to $\sigma_{\langle F \rangle}/\langle F \rangle = 0.8$ for Ly$\alpha$ 
fluctuations. If instead $\langle F_{\alpha} \rangle =0.001$ at $z=6$ (and 
thus $\avg{F_\beta} [z=6] = 0.085$), we obtain $\sigma_{\rm \avg{F_{tot}}}/\avg{F_{\rm tot}} \sim 0.45$, 
compared to $\sigma_{\langle F \rangle}/\langle F \rangle \sim 2$ for Ly$\alpha$ fluctuations. 
Thus, in principle the Ly$\beta$ forest is a much more robust indicator of a fluctuating radiation field. 
Unfortunately, due to contamination from the Ly$\gamma$ forest, at present the Ly$\beta$ forest 
suffers from 
small-number statistics. For instance, Fan et al. (2005) have 97 vs. 19 measurements of 
the Ly$\alpha$ and Ly$\beta$ forests respectively, of size $\Delta z = 0.15$, 
spanning $z= 4.8 - 6.3$ in Ly$\alpha$, and $z = 5.3 - 6.3$ in Ly$\beta$. The observed 
scatter in the Ly$\beta$ forest (their Fig. 3) is consistent with density fluctuations alone.   

\subsection{Sightline-to-Sightline Scatter in the HeII Forest}
\label{heii}

Finally, we point out that a very similar calculation applies to the sightline-to-sightline scatter in
the transmissivity of the HeII Ly$\alpha$ forest near $z \sim 3$. Specifically, we measure the flux
power spectrum in the HeII Ly$\alpha$ forest using a lower redshift simulation output,  
adopting $\tau_{\rm eff, HeII} \sim 4$ at $z = 2.9$ based
on the measurements of Smette et al. (2002). These measurements are performed in 
narrow bins with co-moving length
scales of $L \sim 18 - 36$ Mpc/$h$. At $z = 2.9$, we find that the rms fractional fluctuation in the 
HeII Ly$\alpha$ transmissivity on these smoothing scales 
is $\sigma_{\rm \avg{F, HeII}}/\avg{F_{\rm HeII}} = 1.6 - 1.2$.  
Again, we caution against interpreting the observed scatter as a signature of fluctuations
in the HeII ionizing background.

\section{Conclusions}
\label{conclusion}

We do not mean to imply that the radiation field is indeed uniform at $z \sim 6$.
Our intent is only to show that ruling out the {\em null hypothesis} that
the scatter in the high redshift Ly$\alpha$ forest results solely from density fluctuations is subtle.
Indeed, we find that transmissivity fluctuations should be {\em of order unity} at $z \sim 6$
on scales of $L \sim 50$ Mpc/$h$, from density fluctuations alone. We therefore caution against
over-interpreting the large scatter in $\tau_{\rm eff}$ seen in the spectra 
of $z \sim 6$ quasars or the analogous scatter seen in the $z \sim 3$ HeII Ly$\alpha$ forest.
In future work, we intend to model the effect of
inhomogeneous reionization on the statistics of the Ly$\alpha$ forest, following up on the work of
Furlanetto \& Oh (2005), and to design statistical measures to discern the presence or absence
of these fluctuations.

\section*{Acknowledgments} AL thanks Katrin Heitmann and Salman
Habib for their collaboration in producing the HPM simulation used in
this analysis. We thank Matias Zaldarriaga for useful discussions and comments on a draft.
We thank Lars Hernquist, Lam Hui and Oliver Zahn for stimulating conversations. SPO 
acknowledges NSF grant AST0407084 for support.

%%%%%%%%%%%%%%%%%%%%%%%%%%%%%%%%%%%%%%%%%%%%%%%%%%%%%%%%%%%%%%%%%%%%%%%%%%%

%%%%%%%%%%%%%%%%%%%%%%%%%%%%%%%%%%%%%%%%%%%%%%%%%%%%%%%%%%%%%%%%%%%%%%%%%%%

\end{document}